\documentclass[12pt]{revtex4}
\usepackage{graphicx}
\usepackage{xcolor}

\begin{document}
	
	\title[Quantum matter and gravitation: photons in a waveguide]{Quantum matter and gravitation: photons in a waveguide}
	
\author{Victor Atanasov}	
\affiliation{Department of Condensed Matter Physics and Microelectronics, Faculty of Physics, Sofia University, 5 blvd J. Bourchier, Sofia 1164, Bulgaria}
\email{vatanaso@phys.uni-sofia.bg}

\author{Avadh Saxena}
\affiliation{Theoretical Division and Center for Nonlinear Studies, Los Alamos National Laboratory, Los Alamos, NM 87545 USA}

	\begin{abstract}
		The conditions required by quantum matter to modify space-time geometry are explored within the framework of the general theory of relativity. The required characteristics for space-time modification in solid state structures, are met in either (a) massive photon Bose-Einstein condensate in a waveguide, or (b) the massive photons in superconductor's bulk, or (c) the Bose-Einstein condensate of acoustic phonons, or (d) a metal-insulator-topological insulator heterostructure. 
	\end{abstract}

\maketitle

\section{Introduction}

Gravitational physics remains in its primordial theoretical and observational state, besides the recent spectacular success at gravitational wave detection and the opening up of gravitational wave astronomy \cite{ligo}. Feynman's legacy, namely ``What I cannot create I do not understand'', regarding {\it curvature modification} remains largely unexplored. Here we discuss the few opportunities quantum matter may provide in circumventing the smallness of the Einstein coupling constant $\kappa=8\pi G/c^4$ $\approx 2 \times 10^{-43}$ [m.J$^{-1}$] between the matter content and the curvature of space-time.

In this paper we identify classes of quasi-particulate quantum matter with key main ingredients, namely {\it linear dispersion} and {\it subluminal velocity}, which hold potential to affect the curvature of the embedding space-time. The right solid state and condensed matter contexts now include not only graphene  but a myriad of Dirac quantum materials, such as topological insulators \cite{top1, top2}, where linear dispersion arises as a result of the strong spin-orbit coupling, i.e. in heavy-element small-bandgap semiconductors such as 3D bismuth chalcogenides (Bi$_{1-x}$Sb$_{x}$). The non-trivial aspects of topological insulators are further amplified via the proximity effect when in contact with a superconductor \cite{top3}. The gravitational connection is subtle and has been explored in \cite{top4, top5, top6, top7}. 

Chiral fermions occurring as collective excitations in Weyl semimetal materials \cite{weyl1, top5, weyl2} are known for their anomalous  chiral conservation law which is endowed with new terms generated by the curved and twisted space-time in which the chiral fermions reside. Notably, the mixed axial-gravitational anomaly has already been detected in NbP \cite{weyl3}. The anomaly's origin remains unsettled as to what is the main cause besides the breakdown of classical symmetry at the quantum level, either  (i) it is tied to thermoelectrical transport in flat space-time \cite{weyl21} or (ii) that the gravitational contribution (4$^{\rm th}$ order in derivatives)  gives rise to a fluid velocity, a geometrical effect. There are other important studies of gravitational anomalies in topological quantum matter \cite{ano1, ano2, ano3, ano4, ano5}. 

While the linear dispersion in these fermionic Dirac materials is a consequence of their band structure, there are other origins for the relativistic low-energy quasi-particle excitations (nodal directions throughout the Brillouin zone) in $d$-wave superconductors \cite{dirac}.

Conversely, circumventing the restrictions of the Pauli exclusion principle on particle concentration, there are bosonic Dirac materials \cite{bosedirac} as well as plasmons \cite{bosedirac2}, magnons \cite{bosedirac3}, granular superconductors \cite{bosedirac4}, acoustic phonons \cite{kittel}, including those in liquid superfluid helium and {\it massive} photons, that are a result of a spontaneous symmetry breaking (Anderson-Higgs mechanism) in the electrodynamics of the superconducting state which allows for the Meissner effect \cite{higgs}. 

\bigskip

First, let us take the trace of the $\Lambda$-free Einstein field equations to obtain an average estimate on how matter (including quantum) affects the geometry of space-time:
\begin{eqnarray}
R=-\kappa T.
\end{eqnarray}
Here $R$ is the Ricci scalar curvature and $T$ is the trace of the stress-energy tensor associated with the material content \cite{ein}.

In order to proceed, the simplifying argument requires that we assume the material content being gas-like, i.e. fluid phase, and $T=3p-\rho c^2,$ where $\rho$ is the density, $p$ the pressure of a homogeneous and isotropic phase and $c$ the speed of light in vacuum. Next, let us employ a result from kinetic molecular theory of gases
$p=m n \langle v^2 \rangle /3,$ where $m$ is the mass of the individual identical particles, $n$ is their concentration and $\langle v^2 \rangle$ is the average of their squared velocity \cite{stat}.

In light of the above definitions, the density can be expressed as $\rho=m n.$ Inserting these results in the field equation for the geometry of space-time we obtain
\begin{eqnarray}\label{Rv2/c2}
R=\kappa m c^2 n \left( 1 - \frac{\langle v^2 \rangle}{c^2}  \right) 
\end{eqnarray}
only to ask the question: {\it Is there matter, including quantum, that can affect the gravitational field at the lab scale concentrations?}

To answer the question, we will estimate few of the variables in the formula for the space-time curvature.

\section{Theoretical Analysis}

\subsection{Squared velocity as a quantum mechanical average}

The mean of a velocity component $v_i$ is given by
\begin{eqnarray}\label{}
\langle v_i \rangle = \langle \psi | v_i | \psi \rangle = \langle \frac{d x_i}{dt} \rangle = \frac{1}{i \hbar} \langle [ x_i, H ] \rangle
\end{eqnarray}
which for a free particle $H=H_0=p^2/2m$ is reduced to
\begin{eqnarray}\label{}
\langle v_i \rangle = \frac{\langle p_i \rangle}{m} 
\end{eqnarray}
provided the standard commutational relations hold $[x_i, p_j] = i \hbar \delta_{ij}.$
Now, the squared velocity can be evaluated from
\begin{eqnarray}\label{}
\langle v^2 \rangle = \sum_i \langle \frac{d x_i}{dt} \rangle \langle\frac{d x_i}{dt} \rangle = \frac{1}{(i \hbar)^2} \sum_i \langle [ x_i, H ] \rangle \langle [ x_i, H ] \rangle.
\end{eqnarray}
Here we would assume a composite hamiltonian which includes interaction:
\begin{eqnarray}\label{}
H=H_0+H_{\rm int}
\end{eqnarray}
and renders 
\begin{eqnarray}\label{}
\langle v^2 \rangle &=& \sum_i \left\{ \frac{\langle p_i \rangle \langle p_i \rangle}{m^2} + \frac{2 \langle p_i \rangle}{i \hbar m} \langle [ x_i, H_{\rm int} ] \rangle  + \frac{1}{(i \hbar)^2} \langle [ x_i, H_{\rm int} ] \rangle \langle [ x_i, H_{\rm int} ] \rangle \right\}
\end{eqnarray}
The commutators in the expression can be evaluated provided some piece of information on the dependence of the interaction is known: i.) $H_{\rm int} = H_{\rm int} (x_j)$ is a function of the coordinates only, then $[ x_i, H_{\rm int} ] = 0$ and
\begin{eqnarray}\label{}
\frac{\langle v^2 \rangle}{c^2} = \sum_i \frac{\langle c p_i \rangle \langle c p_i \rangle}{(mc^2)^2} < 1 
\end{eqnarray}
or ii.) $H_{\rm int} = H_{\rm int} (p_j)$ is a function of the momenta, then $[ x_i, H_{\rm int} ] = i \hbar {\partial H_{\rm int}}/{\partial p_i}$ and
\begin{eqnarray}\label{}
 \frac{\langle v^2 \rangle}{c^2} &=& \sum_i \left\{ \frac{\langle c p_i \rangle \langle c p_i \rangle}{(mc^2)^2} + \frac{2 \langle p_i \rangle}{mc^2} \langle \frac{\partial H_{\rm int}}{\partial p_i} \rangle  +  \langle \frac{\partial H_{\rm int}}{\partial c p_i} \rangle \langle \frac{\partial H_{\rm int}}{\partial c p_i} \rangle  \right\}  < 1
\end{eqnarray}
is less than 1 in the non-relativistic case. Therefore, when exploring the gravitational properties of Dirac quantum matter as governed by (\ref{Rv2/c2}), we can conclude that  
\begin{eqnarray}\label{v^2/c^2}
0 < 1 - \frac{\langle v^2 \rangle}{c^2} < 1
\end{eqnarray}
and $R \propto m$ where $m$ is the effective mass of the material content when the velocities are non-relativistic.

Our first observation is a condition on the average squared velocity of these particles, which is $\langle v^2 \rangle < c^2$ and immediately excludes the free photon gas. Besides, for the free photon gas $T=0$ due to the electromagnetic stress-energy tensor being traceless. Therefore, {\it free photon gas cannot modify the curvature of space-time.}

\subsection{Gravitational properties of the effective mass}

The issue whether the effective mass of a quasi-particle is associated with its gravitational response is somewhat similar to the issue whether the mass of a nucleus is associated with its gravitational response, or it is the mass, which is a sum of the bare masses of the constituent nucleons. As is well known, the mass of the nucleus is less than the sum of the masses of the individual nucleons due to the mass defect, i.e. the nuclear binding energy in a quantum mechanical setting leads to an "effective" mass of the nucleus, which is participating in inertial/gravitational interactions. Similarly, the effective mass of quasi-particles, regardless the lower energy scale, is associated with their inertial properties, modified by the periodicity in the solid state context, that is periodicity introduces a restriction on the form of the physically allowed solutions. Indeed, the derivative with respect to time of the quasi-momentum equals the electromagnetic Lorentz force\cite{kittel} for a charged particle in a crystal. The question whether electrons in metals participate in inertial phenomena with their bare or effective mass can be resolved by a careful look at the results behind the Stewart–Tolman effect, which point to the effective mass, rather than the bare, participating in inertial phenomena\cite{ST}. In a conductor undergoing acceleration (incl. centripetal), inertia causes a lag of the electrons, therefore a charge accumulates at the end of the body. The accumulated charge is proportional to the mass of the charge carriers. Moreover, the Stewart–Tolman effect is material specific, which is probably the clearest confirmation for the effective mass of a quasi-particle taking part in an inertial response. We list $m/e$ in units of [10$^{-9}$g/C] obtained in different ways: i.) 5.66 (cathode rays; vacuum); ii.) 5.18 (copper); iii.) 6.73 (silver); iv.) 6.50 (aluminum), etc.

Next, we invoke the Equivalence principle, which in its weak form states that the inertial and gravitational mass coincide, a statement empirically confirmed with a precision of 10$^{-15}$ \cite{microscope}. Therefore, since the effective mass participates in inertial phenomena, it should gravitate as well.

Let us now look at the electron cyclotron resonance in condensed matter physics. This is another instance of the effective mass participating in inertial response, instead of the bare mass. Subject to a static and uniform magnetic field, the electron moves in a circular orbit due to the Lorentz force with a circular frequency of $\omega=eB/m,$ where $e$ is the electron's charge (invariant), $B$ is the magnetic field strength and $m$ is the effective mass. The electron cyclotron resonance results for various materials clearly speak of an effective mass taking part in the inertial response to the applied magnetic field.

\bigskip

The profound difficulty with Ernst Mach idea of mass as a measure of inertia, i.e. resistance to acceleration is its failure to account for the interaction (mass defect) when masses are brought together for measurement. Here the nuclear mass serves as a quintessential example of why this definition of mass is best useful when comparing masses, that is $m_1/m_2 \propto a_1/a_2,$ which is analogous to the approach taken by Coulomb when he justified the necessity to introduce the electric charge. 

Next comes the challenge brought by the impossibility to measure the instantaneous acceleration, i.e. to measure acceleration (therefore mass) with a single measurement. Instead, one takes multiple measurements (position \& time) and does a computation to obtain the acceleration. As a result, the idea that mass is a measure of resistance to acceleration is flawed by being a non-instantaneous quantity. It is also a non-local quantity. The weakness is further amplified when we consider curved space-time background, where the computational procedure requires  values belonging to different tangent spaces and mass becomes an ambiguous quantity, if not meaningless.

Therefore, we assume that the quasi-classical quantum interpretation of mass, which is common throughout the solid state or condensed matter \cite{kittel}, as a function of the curvature of the dispersion relation (energy band) is superior to Mach's definition in i.) taking into account the interaction encoded in $\mathcal{E}(\vec{k})$; ii.) being instantaneous for the stationary quantum mechanical case and iii.) being non-local in momentum space, instead of coordinate space. 

Here $\mathcal{E}(\vec{k})$ is the dispersion relation, i.e. the energy as a function of the wavevector ($\vec{p}=\hbar \vec{k}$ momentum). The variable
\begin{eqnarray}\label{quasiclassM}
m=\frac{\hbar^2}{\mathcal{E}^{\prime \prime} }  
\end{eqnarray}
is the effective mass of the quasi-particle and generally is of tensorial character. Here $\mathcal{E}^{\prime \prime}=\partial^2 \mathcal{E}/\partial k^2 $ and $\hbar$ is the reduced Planck constant.

We are faced with an ambiguity related to the definition of mass within the condensed matter physical models, which ambiguity is usually resolved by switching between models and definitions. For example, when considering the dynamics of particles with linear dispersion, we take the Klein-Gordon or Dirac equations and postulate that the bare mass is the limit to the dispersion relation when momentum vanishes (here $p=\hbar k$):
\begin{eqnarray}\label{rel_disp}
m_0 =\lim_{k \to 0} \; \frac{1}{c^2}\sqrt{(m_0 c^2)^2 + \hbar^2 c^2 k^2}  
\end{eqnarray}

Therefore, we have a bare mass definition of the type
\begin{eqnarray}
m_0 =\lim_{k \to 0} \; \frac{1}{c^2} \mathcal{E}(k) , 
\end{eqnarray}
which is practically vanishing for any solid state or condensed matter setting.

It is interesting to see for what dispersion relation the effective mass coincides with the bare one. For the purpose we solve the following differential equation $\mathcal{E}^{\prime \prime}={\hbar^2}/{ m_0}$. The solution is $
\mathcal{E}(k)= \frac{\hbar^2 k^2}{2 m_0} + \mathcal{E}_0
$ where we make the observation that the effective mass coincides with the bare one only for free motion (no interactions) with non-relativistic velocities. Therefore, {\it we cannot expect the effective mass to coincide with the bare one in any interacting system.}

The relativistic case reveals something peculiar, namely in the limit $k \to 0$ both the bare and the effective masses coincide. For the relativistic dispersion law in (\ref{rel_disp}) we obtain:
\begin{eqnarray}\label{}
m = \frac{\hbar^2}{\mathcal{E}^{\prime \prime} }=
m_0 \left( 1 + \frac{c^2 \hbar^2  k^2}{m_0^2 c^4 }   \right)^{\frac{3}{2} }
\end{eqnarray}
therefore for small momenta 
\begin{eqnarray}
\lim_{k \to 0} m_0 \left( 1 + \frac{c^2 \hbar^2  k^2}{m_0^2 c^4 }   \right)^{\frac{3}{2} } = m_0 
\end{eqnarray}
and in the relativistic case the expression for the effective mass coincides with the one for the bare mass.

In the case of the optical effective mass, that is an energy and momentum dependent definition of the effective mass\cite{ariel}:
\begin{eqnarray}\label{}
m=\frac{\hbar^2 k}{\mathcal{E}^{\prime} } = \frac{\hbar^2}{2 \pi} \frac{\partial A}{\partial E},  
\end{eqnarray}
is formally equivalent to the cyclotron effective mass \cite{shokley}. Here $A(E)$ is the area bound by the closed path in momentum space travelled by an electron at an energy level $E.$ For the relativistic dispersion law in (\ref{rel_disp}) we obtain for the optical effective mass $m=p/v$:
\begin{eqnarray}
m= \frac{\hbar^2 k}{\mathcal{E}^{\prime} } = m_0 \sqrt{1 + \frac{c^2 \hbar^2  k^2}{m_0^2 c^4 } }  , 
\end{eqnarray}
which limit in the case of vanishing momenta coincides with the bare mass once again.
\begin{eqnarray}\label{equivalentDEF}
\lim_{k \to 0} \frac{\hbar^2}{\mathcal{E}^{\prime \prime} } 
= \lim_{k \to 0} \; \frac{\mathcal{E}}{c^2}
=\lim_{k \to 0} \frac{\hbar^2 k}{\mathcal{E}^{\prime} }
=m_0
\end{eqnarray}

Presently, we have no way to choose a proper definition of the effective mass to use in our gravitational exploits. For that choice we can only be guided by the Equivalence principle, namely if a kind of mass resists accelerations, i.e. it is inertial, then it is gravitational as well. Note, for dispersion relations of the type (\ref{rel_disp}) all definitions of the effective mass restore the same bare mass limit (see eq. (\ref{equivalentDEF})). Therefore, we can pick the definition (\ref{quasiclassM}) as the proper definition for a quasi-particle mass involved in gravitational interaction. It describes non-relativistic quantum matter, that is one that has non-vanishing effective mass and as a result (\ref{v^2/c^2}) holds for the benefit of non-vanishing gravitational effect.

In this case, we can also write
\begin{eqnarray}\label{vgr}
\vec{v}_{gr}=\frac{\partial \omega}{\partial \vec{k}}=\frac{1}{\hbar}\frac{\partial \mathcal{E}}{\partial \vec{k}}
\end{eqnarray}
as the group velocity of quasi-particle's wave packet.

We would like to end the section by making the distiction between quasi-particles and '''quasi-particles'', where the former being  I.) particles which exist in vacuum and can have their dispersion laws modified in solid state structures, i.e. become effectively quasi-particles (photon in a waveguide, protons \& neutrons bound in a nucleus, ect.) and; the latter II.) quasi-particles which exist in a material medium (like polarons, phonons, Dirac/Majorana fermions, ect.) and are non-existent by themselves in vacuum. In this paper we will deal with quasi-particles of the type I.

\subsection{The solution}

Inserting (\ref{quasiclassM}) and (\ref{vgr}) into (\ref{Rv2/c2}) we obtain
\begin{eqnarray}\label{E1}
\frac{\mathcal{E}^{\prime} }{\mathcal{E}^{\prime}}  \frac{\mathcal{E}^{\prime \prime} }{1 - \frac{\mathcal{E}^{\prime \; 2}}{\hbar^2 c^2}  }= \frac12 \frac{(\mathcal{E}^{\prime \;2})^\prime }{\mathcal{E}^{\prime} - \frac{\mathcal{E}^{\prime \; 3}}{\hbar^2 c^2}  }=
\frac{\kappa \hbar^2 c^2 n}{R}  
\end{eqnarray}
The substitution with the dimensionless variable
\begin{eqnarray}
\frac{\mathcal{E}^{\prime}}{\hbar c}  = u
\end{eqnarray}
simplifies equation (\ref{E1}):
\begin{eqnarray}
\frac{ u^\prime}{1 - u^2  }=
 \frac{\kappa \hbar c n}{R} 
\end{eqnarray}
which can now be easily solved for the Ricci scalar curvature 
\begin{eqnarray}\label{R_tanh}
R=\frac{ \kappa \hbar c n }{ d \tanh^{-1}{(u)} / dk}
\end{eqnarray}
or for the dispersion law's derivative as a function of the Ricci scalar curvature
\begin{eqnarray}\label{Eprime}
\mathcal{E}^{\prime}=c \hbar\tanh{\left(C_1 +
\kappa \hbar c n \int \frac{d k}{R} \right)}.
\end{eqnarray}

Provided the following holds
\begin{eqnarray}
\frac{d \tanh^{-1}{(u)} }{ dk} =0 
\end{eqnarray}
i.e.
\begin{eqnarray}\label{rel_disp}
\mathcal{E} (k)= \alpha c \, \hbar k + \mathcal{E}_0,
\end{eqnarray}
where $\alpha$ and $\mathcal{E}_0$ are a constant, the curvature of space-time diverges $R=\infty$ as a consequence of a vanishing denominator in (\ref{R_tanh}), i.e. a potentially measurable effect \cite{lalov}.

The integration constant $\alpha<1,$ since
\begin{eqnarray}
\lim_{u \to 1}\tanh^{-1}{u}=\infty \quad \Rightarrow \; 
\mathcal{E}^{\prime} \to c \hbar \; ; \; R \to 0,
\end{eqnarray}
that is, if the dispersion relation is linear and the velocity of propagation is the velocity of light in vacuum, then the curvature of space-time is vanishing. Freely propagating photons do not produce curvature of space-time. They merely follow geodesic trajectories.

Note, regardless of how small the coupling constant $\gamma=\kappa  \hbar c n$ is, the denominator in (\ref{R_tanh}) can produce a substantial effect since $R$ is divergent when
\begin{eqnarray}
\mathcal{E}^{\prime \prime} = 0.  
\end{eqnarray}
This is clearly possible for the relativistic dispersion law in (\ref{rel_disp}).

\section{Experimental realizations}

In practice, this is the dispersion law obeyed by the fermionic carriers in Dirac materials \cite{dirac}, specifically graphene \cite{neto}, topological insulators \cite{top1, top2} and a variety of Dirac and Weyl semimetals \cite{weyl1, weyl2}. Bosonic Dirac matter also possesses linear dispersion \cite{bosedirac}. In graphene $v_{gr}$ coincides with the Fermi velocity $\sim 10^6$ [m.s$^{-1}$] and $\langle v^2 \rangle / c^2 \approx 10^{-5}$ which is in accord with the requirement (\ref{v^2/c^2}). Note also that the concentration $n$ is the only experimentally controllable parameter in the numerator of (\ref{R_tanh}) and needs to be nonvanishing.

Here we cannot exclude that the  material specific dispersion relation $\mathcal{E}(\vec{k})$ is of the type $\sim |\vec{k}|,$ which for $\vec{k}=0$ would not have well defined derivatives. As a result the left/right limit can lead to a sign change. The derivatives with respect to various Brillouin zone directions can also lead to distinct $R\to \pm \infty$ limiting procedures.

One obvious disadvantage of the relativistic fermion gas is the Pauli's exclusion principle, which limits the concentration $n$.  However, this limitation is not there for the bosonic Dirac matter \cite{bosedirac}. For graphene $n \sim 10^{17}$ [m $^{-2}$] is purely two-dimensional, which cannot be considered as a nonvanishing three-dimensional concentration in itself. It is a three-dimensional concentration only in the sense of a distribution, which would be an ill-defined source in Einstein field equations \cite{sources}. Therefore, for graphene the numerator in (\ref{R_tanh}) is vanishing and $R=0,$ which necessarily points to the exploration of other material realizations.

In order to circumvent Pauli's exclusion principle, we need to consider a Bose-Einstein condensate of relativistic bosons, for which the concentration $n$ is not limited from above. As already discussed, it cannot be of free photon nature, however acoustic phonons (and magnons in bosonic Dirac matter \cite{bosedirac}) are examples of bosonic quantum particles with almost linear dispersion for low wavenumbers \cite{rossen, kittel}. The acoustic phonon case will be discussed elsewhere, but the speed of sound in the lattice is orders of magnitude smaller than $c$. The acoustic phonon pseudo-relativistic condensate is the reason for long wavelength packets to propagate over large distances across the lattice without breaking apart. Phonons can also be considered to have negative mass, and in effect, have negative gravity \cite{phonon1, phonon2}.

\subsection{Massive photons in superconductors}

Now, let us turn to the massive vector spin-1 boson field recognized as a modification to Maxwell's electrodynamics which allows for photon mass while preserving Lorentz invariance \cite{goldhaber, jackson}. More importantly, in the context of (\ref{Rv2/c2}) and (\ref{R_tanh}), a massive photon field can realize the necessary condition $\langle v^2 \rangle < c^2$ while preserving a quasi-linear dispersion. The consequences resulting from a non-vanishing photon mass, besides the altered dispersion relation (frequency dependent velocity of propagation), are (i) the exponential fall-off (Yukawa potential) of fields with distance; (ii) the appearance of a pseudocharge, i.e. local charge non-conservation \cite{noncons1} and (iii) a violation of gauge invariance, which can be due to an implied additional dynamics, i.e. another field interacting with the photon field. 

The breakdown of gauge symmetry, that is the Anderson-Higgs mechanism \cite{higgs} in the context of the electrodynamic behavior of superconductors produces the Meissner effect: $\vec{j} \propto \vec{A},$ that is the expulsion of the electromagnetic field from the bulk of the superconductor. Microscopically, the perfect diamagnetism arises due to the effective photon mass gained via the interaction of the electromagnetic field with the charged bosonic condensate (Cooper quasi-particle pairs \cite{BCS}). Therefore, a condensed matter context exists (type I or II superconductors), which prompts the realization of a process of an effective photon mass gain in pursuit of space-time curvature response according to (\ref{R_tanh}).

 The wave equation for the 4-vector potential ${\bf A}$ (i.e. photon wavefunction) in the superconductor takes the form
\begin{eqnarray}
\left(\nabla^2 - \partial^2_{c t} - \frac{1}{\lambda^2_{L}} \right) {\bf A} =0,
\end{eqnarray}
where $\lambda_{L}$ is the London penetration depth ($\sim$ several tens of nm). The resulting dispersion relation is
\begin{eqnarray}\label{E_k}
\mathcal{E}(k)=\pm c \hbar k\sqrt{1+\left( \frac{1}{k\lambda_L} \right)^2} , 
\end{eqnarray}
which for $k\lambda_L \gg 1$ can be approximated by
\begin{eqnarray}
\mathcal{E}(k)=\pm c \hbar k, 
\end{eqnarray}
i.e. a relativistic dispersion, while the group velocity
\begin{eqnarray}\label{v_k}
v_{gr}=\frac{d\omega}{d k}=c/\sqrt{1+\left( \frac{1}{k\lambda_L} \right)^2} < c
\end{eqnarray}
is less than $c$ and frequency dependent. Therefore, for the massive photon in superconductors the restriction $\langle v^2 \rangle < c^2$ holds. The effective mass of the photon in superconducting materials $m=\hbar/\lambda_L c$ is of the order of few eV. Naturally, the massive photon is a boson and undergoes Bose-Einstein condensation for which $n \gg 1,$ i.e. the concentration, can be arbitrarily large thus fixing the numerator of (\ref{R_tanh}).

Unfortunately, the Meissner effect is a surface phenomenon, that is the superconducting charge density, ergo massive photon field $\vec{j} \propto \vec{A}$, exists only at a $\lambda_L$ skin deep surface layer. In the bulk of the superconductor $\vec{A}=0$. Therefore the necessary concentration of such a massive photon to bend space-time might not be possible to achieve.

\subsection{Massive photons in waveguides}

The photon (and any quantum quasi-particle) is an elementary excitation of the fields. Therefore, the transition of the photon from one environment into another completely changes its properties. Any material environment is characterized by its index of refraction  ${\rm n}=\sqrt{\epsilon_r \mu_r}$, which is also a measure of how much the photon is slowed down by the interaction with the medium. As a result, one can perceive the photon in a material medium as a quasi-particle with relativistic dispersion law and a velocity of propagation $v=c/{\rm n} < c,$ which are the requirements for curvature modification according to (\ref{R_tanh}). 

Therefore, we suggest solid state structures which serve the purpose of creating an artificial environment which slows the photon down while preserving its linear dispersion. Naturally, there are multiple ways to achieve that and we will discuss i.) the photon interacting with the superconducting pairs and ii.) the photon being spatio-geometrically constrained in a narrow waveguide, which turns out to be equivalent to i.).

\subsubsection{Superconducting waveguides}

\begin{figure}[t]
\begin{center}
\includegraphics[scale=0.4]{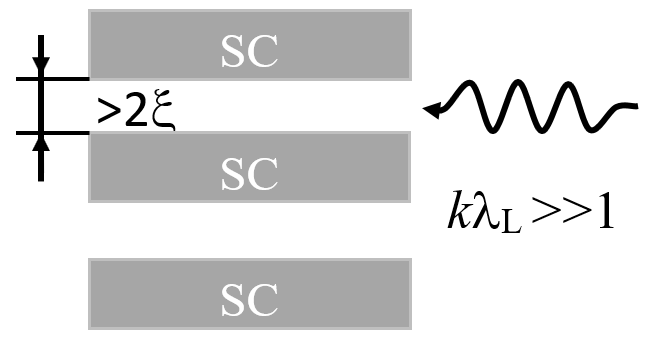}
\caption{\label{fig:SCwave} A layered superconducting structure (with spacer layers in between) acting as a waveguide accumulating massive photon vector bosons with extremely small wavelengths $\lambda \ll 2 \pi \lambda_L \sim 10^{-7}$m. An additional carrier frequency in the THz range can resonantly increase acoustic phonon concentration thus providing an auxiliary mechanism to modify the Ricci scalar curvature $R$ as well as to augment the onset of superconductivity.}
\end{center}
\end{figure}

Thus, an experimental approach to test the massive photon-gravitational interaction emerges. It rests on (i) the proximity effect in superconductors, namely a thin enough film of non-superconducting material placed onto a superconductor acquires superconducting properties due to the spread of Cooper pair wavefunction over distances on the order of the coherence length $\xi$ ($\sim$ several hundreds of nm to $\mu$m) and (ii) waveguide cut-off frequency which ensures that electromagnetic waves of sufficiently large wavenumbers $k \lambda_L \gg 1$ penetrate the layered structure depicted on Figure \ref{fig:SCwave}. The condition is necessary in order to guarantee a proper relativistic approximation to the dispersion law (\ref{E_k}) in the waveguide. 

Since $\lambda_L \sim 10^{-8}$m, the wavelength of the corresponding electromagnetic radiation should be $\lambda \ll 2 \pi \lambda_L \sim 10^{-7}$m, which is in the X- or $\gamma$- ray range, i.e. highly energetic radiation.

This structure acts as an accumulator for massive photons and consists of alternating layers of superconducting material, separated by thin (at least twice the coherence length $\xi$) spacer layers.

Here an auxiliary process might become resonant in the few THz range of an additional carrier frequency with an overall effect of an acoustic phonon population increase, subsequent condensation and provision of a secondary mechanism to modify $R$ according to (\ref{R_tanh}). THz pulses can resonantly drive selected vibrational modes in solids, deform crystal structures, melt electronic order, drive insulator-to-metal transitions and induce superconductivity \cite{eliashberg, e2, e3, e4, e5, e6}. The prospect of manipulating the superconducting properties of the structure with the electromagnetic radiation that is being condensed in the waveguide (provided the spacing layers are dielectric) has the following feature: $\lambda_L \propto 1/\sqrt{n_s},$ where $n_s$ is the density of the superconducting carriers in the material. Therefore, the effective mass of the photon $m \propto \sqrt{n_s}$ becomes a frequency controllable parameter.  

The presence of superconducting material introduces an additional requirement $\vec{A}\cdot\vec{n}=0,$ where $\vec{A}$ is the vector potential and $\vec{n}=0,$ is the normal to the superconducting surface. This condition leads to $\vec{B} \times \vec{n}=0,$ therefore TE terahertz modes are not allowed in the waveguide and only TM modes can propagate. Additionally, due to the long wavelengths of THz radiation, the waveguides should have at least one of their lateral dimensions in the mm range, i.e. macroscopically sized widths.

As a natural extension to the properties stemming from the linear dispersion in various Dirac materials, the layered structure allows for variations in its composition such as the waveguides being deposited from topological insulators, Weyl semimetals, honeycomb ferromagnets (exhibiting Dirac magnons), etc.

\subsubsection{X- or $\gamma$- ray waveguides}

X- or $\gamma$- ray waveguides are multilayer
structures with alternating layers of low-density and
high-density materials\cite{gamma}. Accumulation of massive X- or $\gamma$- ray radiation might be possible, because at these photon energies denser materials generally have lower refractive indices than less dense materials and in effect the radiation can be confined  within the condensed matter structure.

The dispersion relation for photons in rectangular waveguide is
\begin{eqnarray}
\mathcal{E}(k)=\hbar \omega=\hbar v \sqrt{k^2 + \left( \frac{{\rm n}_1 \pi }{a}\right)^2  + \left( \frac{{\rm n}_2 \pi }{b}\right)^2}
\end{eqnarray}
Here the phase velocity is $v$ and is material specific and ${\rm n}_i \in \mathcal{Z}$ for $i=1,2$. The confining dimensions of the waveguide are denoted with $a$ and $b$.  The mass of the photons in such a waveguide can be calculated according to (\ref{quasiclassM})
\begin{eqnarray}
m=\frac{\hbar} {v}   \frac{ \left\{k^2 + \left( \frac{{\rm n}_1 \pi }{a}\right)^2  + \left( \frac{{\rm n}_2 \pi }{b}\right)^2 \right\}^{3/2} }{\left( \frac{{\rm n}_1 \pi }{a}\right)^2  + \left( \frac{{\rm n}_2 \pi }{b}\right)^2 }
\end{eqnarray}
and exhibits wavelength dependence in addition to the dependence on the confining dimensions of the waveguide. 

Here the dimensions of the waveguide play the same role as the London penetration depth $\lambda_L$ in (\ref{E_k}) and (\ref{v_k}) and these results hold equivalently provided $k a \gg 1,$ where $a<b.$ Since $a$ is of the order of $\mu$m (thickness of the spacer layers in a narrow waveguide), then the estimate based on $\lambda_L$ for the required wavelengths holds in that case as well.

The Ricci scalar curvature induced by the confined photons is 
\begin{eqnarray*}\label{}
R=\kappa \hbar c n |k|^3 \frac{  \left\{1 + \left( \frac{{\rm n}_1 \pi }{ak}\right)^2  + \left( \frac{{\rm n}_2 \pi }{bk}\right)^2 \right\}^{3/2} \left( 1 - \frac{1}{|{\rm n}|^2}  \right)  }{  \left( \frac{{\rm n}_1 \pi }{a}\right)^2  + \left( \frac{{\rm n}_2 \pi }{b}\right)^2  }   {\rm n}
\end{eqnarray*}
Here ${\rm n}=c/v$ is the index of refraction and $k=2 \pi {\rm n}/\lambda_0,$ where $\lambda_0$ is the wavelength in vacuum. 

In the limit $ak \gg 1$ and $a \ll b$, that is a layered structure for which the lateral size is much bigger than the thickness (carpet-like layers) 
\begin{eqnarray}\label{}
R=\kappa \hbar c\;  n  \frac{ 8 \pi  a^2  }{  {\rm n}_1^2 \lambda_0^3}  \left( 1 - \frac{1}{|{\rm n}|^2}  \right) |{\rm n}|^3 {\rm n}
\end{eqnarray}
the expression is largely simplified. Indeed the produced curvature is far from being substantial, but is non-vanishing. Here $\kappa \hbar c \approx 6 \times 10^{-69}$ [m$^2$] and is unlikely to be substantially overcome by a factor of the type
\begin{eqnarray}\label{}
\frac{\rm concentration \times size^2}{\rm wavelength^3}.
\end{eqnarray}

Suppose the waveguide's dimension $a \sim$ 100 $\mu$m and $n \sim$ 10$^{28}$ [m$^{-3}$], then for $\gamma$-rays from a collimated $^{57}$Co source (122 keV or $\lambda=1 \times 10^{-11}$m ) we obtain the estimate $n  \frac{ 8 \pi  a^2  }{  {\rm n}_1^2 \lambda_0^3} \sim 2 \times 10^{53}$ [m$^{-4}$], where ${\rm n}_1=1$ mode. While the dimensions of the waveguide might be fixed, the concentration of the photons inside and their wavelength are variable experimental parameters. Note, that the concentration and wavelength determine the energy density of the photon field inside the waveguide. In the above case $w_{EM} = n \hbar \omega \approx$ 10$^{14}$  [J . m$^{-3}$] . Note, from electrodynamics we have 
$w_{EM} = \epsilon_0 E^2$, where $\epsilon_0$=8.85 x 10$^{-12}$ F/m , which points to $E$ being of the order of 10$^{12}$ V/m. 

Next, suppose the index of refraction ${\rm n} \sim 1.5$ then the contribution from the last factor is of the order of unity. However
\begin{eqnarray}\label{}
R \propto {\rm n}=\pm  |{\rm n}|,
\end{eqnarray}
that is the sign of the space-time curvature depends on the sign of the index of refraction which can take negative values for meta-material layered structures. Such a sign inversion is not possible when considering inertial masses and their one type only $m>0$ gravitational properties. A meta-material waveguide layered structure can exhibit $m<0$ negative mass like gravitational response.

Since the available materials vary in density, we may demonstrate massive photon accumulation in a fabricated multilayer structure consisting of 20-30 alternating layers of few tens of $\mu$m thick sputter or evaporation deposited Pb and few $\mu$m thick spin-coated UV curable polymer film. Here Pb is the dense and durable material, while the organic polymer spacers will undergo degradation during the operation. Such a waveguide can be readily manufactured from thin lead foil with an applied oil film on it. Stacking such oiled layers may produce the desired waveguide structure in the first approximation. The oil's organic molecules will provide for the spacing between the dense Pb layers.

\begin{figure}[t]
\begin{center}
\includegraphics[scale=0.4]{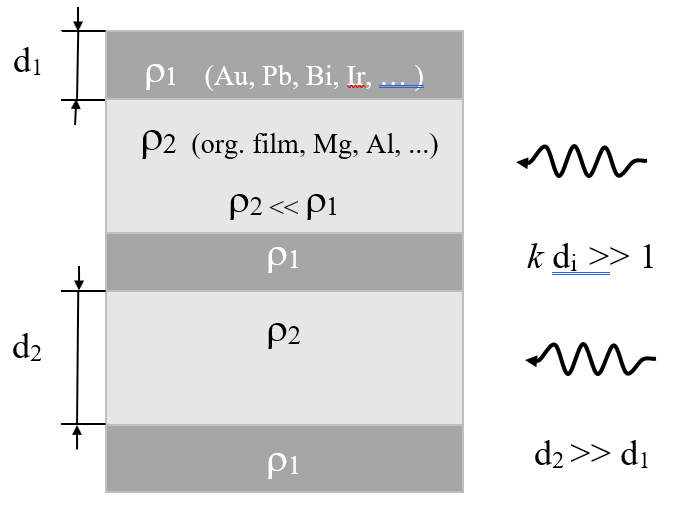}
\caption{\label{fig:wg} An X- or $\gamma$- ray waveguide is a multilayer
structure with alternating layers of low-density (organic film, Mg, Al, etc.) and
high-density materials (Bi, Pb, Au, Ir, etc.). The waveguide can take the form of closed shell for the purpose of massive photon accumulation. Another realization inspired by the optical properties of metals suggests the use of high-conductivity (Cu, Al, Ag, etc.) and low-conductivity (Nichrome , graphite) materials for the layers.
}
\end{center}
\end{figure}

 A second similar structure consists of 20-30 pairs of few $\mu$m thick Bi or Pb layers and few tens of $\mu$m thick Al or Mg layers. Here the bismuth layers serve as reflectors (denser substitutes include U, Pt and Ir) while the low-density aluminum or magnesium layers serve as spacers where massive photons abide. This second manufacturing proposition is a durable structure and should not undergo decomposition during operation. A visual depiction is given on Figure \ref{fig:wg}.

Next, $\gamma$-rays from a collimated $^{57}$Co source (122 keV or $\lambda=1 \times 10^{-11}$m ) or other source strike the edge of the multilayer waveguide to provide for the massive photon Bose-Einstein condensate accumulation. For the purpose of facilitation of the accumulation, the waveguide can be manufactured with a closed shell geometry. A closed shell topology with a smoothly varying curvature can contain the radiation in the shell and which weight (reaction to the support) can be measured as a way of verifying the space-time geometrical modification. The closed shell topology allows for an increase in the  concentration $n,$ which would be more difficult in the open shell configuration and would require intense radioactive source.  Losses can be reduced by the reduction of the scattering cross-section and by flattening the structure as much as possible (see Klein–Nishina formula\cite{weinberg}). 

Note, the design and material choice for a $\gamma$-ray waveguide can go through purely optical consideration where $\gamma$-rays interact with an effectively free electron gas. In this case, the dielectric permittivity of metals model suggests the alternate use of high (Cu, Al, Ag, etc.) and low (Nichrome , graphite) conductivity materials for the layers. An example includes Al/graphite multilayer structure.

\subsubsection{Fermionic case}

\begin{figure}[t]
\begin{center}
\includegraphics[scale=0.4]{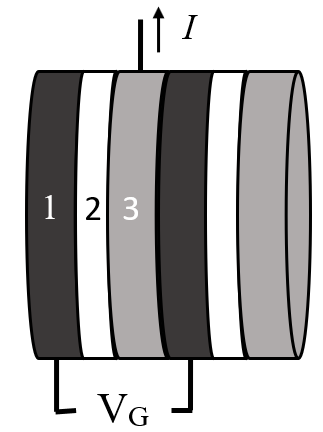}
\caption{\label{fig3} A possible 1-2-3 superlattice structure for the exploration of curvature modification with Dirac fermionic material. Here 1 is bismuth, 2 is an insulator (Bi$_2$O$_3$) and 3 is a topological insulator (bismuth chalcogenide). The applied gate voltage $V_G$ provides tunable relativistic mass and topological transport regime for draining the current $I\propto n,$ which is proportional to the relativistic carrier concentration. 
}
\end{center}
\end{figure}

The exploration of the fermionic realization of gravitational modification with Dirac matter can be achieved with a layered superlattice which consists of alternating metal-insulator-topological insulator (M-I-TI) layers depicted in Figure \ref{fig3}. 

The necessity for a heterostructure is a product of the impossibility of any fermionic material to exhibit the desired {\it linear dispersion} and {\it subluminal velocity} quasi-particles without tuning the material properties. The tuning goes through the application of the field effect which requires an intermediate insulator layer, hence the sandwich structure of the type conductor-insulator-conductor. The topological insulator is the material where the target quasi-particles emerge but only as a surface phenomenon. As a result, thin film manufacturing techniques should be combined to produce a super-lattice where the discussed curvature modification can take place.

Bismuth based materials have been instrumental in the development of topological physics and bismuth itself has emerged as a higher-order topological insulator and first-order topological crystalline insulator \cite{Bi1, Bi2}. Therefore, the superlattice can be realized with the metal layer being bismuth itself. The insulating layer, instrumental for the gate control of the relativistic mass and topological transport parameters, can be realized with bismuth trioxide Bi$_2$O$_3.$ The topological insulator layer adjacent to the bismuth layer can be composed of a bismuth chalcogenide. To access the topological states in the structure, one tunes the relativistic mass regime by gate-controlling the shift in the chemical potential. Additional control of the carrier density $n$ is possible by carrying current through contacts on the topological insulator layer.

Such a stacked structure can be grown on a support substrate with the initial gate layer being bismuth (chosen due to its being HOTI/TCI; other metals work for gates as well), which thin film deposition is complicated due to its low melting point ($\approx$ 545K) that leads to high crystallization rate and low vapour pressure. The deposition can be carried out with several deposition techniques, including pulsed laser deposition (PLD), rf  sputtering ,  thermal  evaporation, etc. The thickness of the gate layers could be in the few to tens of $\mu$m range. Next comes a  thin $\sim$ 30 nm high-k dielectric (Al$_2$O$_3$, SrTiO$_3$ or the native Bi$_2$O$_3$) on top of the gate. This dielectric allows for  variation in the carrier concentration (several times) and large tuning of the chemical potential. The quintessential layer is the TI one, where the choice naturally falls on the bismuth chalcogenides Bi$_2$Se$_3$ or Bi$_2$Te$_3,$ for which various growth techniques have been developed such as the vapour-liquid-solid (VLS) process ($\sim$ 30-100nm thickness) or the catalyst-free vapour solid method (down to few nm thickness), where the lateral dimension of the TI grows much faster than the vertical thickness dimension and in principle can lead to wide samples\cite{kong}. Ideally, the TI layer should be as thin as possible with large surface-to-volume ratio, which is electrically gated more successfully than the bulk form to enhance the surface state effects, i.e. tune the Fermi energy (EF) so that the Dirac fermion is expressed. Generally, nano-structures with a high specific areas, have been shown to enhance the surface conduction contribution. The structure is then repeated manyfold. In the case of a single field effect transistor like sequence, a top metal gate is placed on the TI layer. Due to the TI layer being in the nm thickness range, side contacts are very unlikely to be placed successfully, and therefore the TI layer should be contacted from the top where the contacts cut (and be isolated from) the top metal gate. The top metal gate is again in the few to tens of $\mu$m thickness range. 

Note, the manufacture of such layered nano-structures with large lateral sizes where, in order to realise the topological transport regime, materials with grossly different growth techniques are involved and is presently a challenge. We also expect that the Pauli exclusion principle would prevent reaching relevant concentrations of the Dirac matter carriers to exhibit noticeable effect.

\section{Conclusions}

In conclusion, we have explored the conditions required by quantum matter to modify space-time geometry 
within the framework of general theory of relativity. The necessary conditions for this modification to take place are: (i) the mean squared velocity $\langle v^2 \rangle$ of quantum particles should be smaller than $c^2$, (ii) the three-dimensional concentration $n>0$ of the quantum particles should be substantial, and (iii) the dispersion law should be relativistic, i.e. $\mathcal{E}^{\prime \prime}=0$.

These requirements hold for the quasi-particles in both fermionic and bosonic Dirac materials and we have suggested possible condensed matter structures where the list of conditions is fulfilled provided they are irradiated with electromagnetic waves with appropriate wave-numbers in order to accumulate massive photon condensate with sufficient concentration. These suggested device configurations open up gravitation for possible experimental exploration via a variety of quantum matter including topological one.

\section*{Acknowledgements}

The authors report no competing financial interests or specific grants received for this work. Author contributions are as follows: V. A. devised and wrote the initial manuscript with valuable inputs, discussion, comments and significant edits from A. S. The work of A. S. at Los Alamos National Laboratory was carried out under the auspices of the U.S. DOE and NNSA under Contract No. DEAC52-06NA25396.

\end{document}